\documentclass[oldversion]{aa}
\usepackage{graphicx,times}
\usepackage{natbib}
\bibpunct{(}{)}{;}{a}{}{,}
\def\mrk{Mrk~279}

\def\ch{{\it Chandra}}

\def\xmm{XMM-{\it Newton}}
\def\epi{Epic-PN}

\def\Halpha{\ifmmode {\rm H}\alpha \else H$\alpha$\fi}
\def\Hbeta{\ifmmode {\rm H}\beta \else H$\beta$\fi}
\def\Hgamma{\ifmmode {\rm H}\gamma \else H$\gamma$\fi}
\def\Hdelta{\ifmmode {\rm H}\delta \else H$\delta$\fi}
\def\Lya{\ifmmode {\rm Ly}\alpha \else Ly$\alpha$\fi}
\def\Lyb{\ifmmode {\rm Ly}\beta \else Ly$\beta$\fi}
\def\Lyg{\ifmmode {\rm Ly}\beta \else Ly$\gamma$\fi}
\def\fei{Fe\,{\sc i}}

\def\ciii{\ifmmode {\rm C}\,{\sc iii} \else C\,{\sc iii}\fi}
\def\civ{\ifmmode {\rm C}\,{\sc iv} \else C\,{\sc iv}\fi}
\def\cv{\ifmmode {\rm C}\,{\sc v} \else C\,{\sc v}\fi}
\def\cvi{\ifmmode {\rm C}\,{\sc vi} \else C\,{\sc vi}\fi}

\def\nvi{N\,{\sc vi}}
\def\nvii{N\,{\sc vii}}

\def\o5007{[O\,{\sc iii}]\,$\lambda5007$}

\def\ov{O\,{\sc v}}
\def\ovi{O\,{\sc vi}}
\def\ovii{O\,{\sc vii}}
\def\oviii{O\,{\sc viii}}

\def\neix{Ne\,{\sc ix}}
\def\nex{Ne\,{\sc x}}

\def\fei{Fe\,{\sc i}}

\def\feix{Fe\,{\sc ix}}
\def\fexvii{Fe\,{\sc xvii}}

\def\fexxv{Fe\,{\sc xxv}}
\def\fexxvi{Fe\,{\sc xxvi}}

\def\o{\o}
\def\kms{{km\,s$^{-1}$}}
\def\fek{{\rm Fe\,K$\alpha$}}
\def\fekb{{\rm Fe\,K$\beta$}}

\begin{document}

\title{\xmm\ unveils the complex iron K$\alpha$ region of Mrk~279}

\author
{E.~Costantini \inst{1}
\and 
J.S.~Kaastra \inst{1,2}
\and 
K.~Korista \inst{3}
\and 
J.~Ebrero \inst{1}
\and 
N.~Arav \inst{4}
\and 
G.~Kriss \inst{5}
\and 
K.C.~Steenbrugge \inst{6}
}

\offprints{E. Costantini}
\mail{e.costantini@sron.nl}

\institute{ 	SRON National Institute for Space Research, Sorbonnelaan 2, 3584\,CA Utrecht, The Netherlands
              	\and
				Astronomical Institute, Utrecht University, PO Box 80000, 3508\,TA Utrecht, The Netherlands
				\and
	      	Department of Physics, Western Michigan University, Kalamazoo, MI 49008, USA
		\and
		Department of Physics, Virginia Tech, Blacksburg, VA 24061, USA
	      	\and	      
             	Space Telescope Science Institute, 3700 San Martin Drive, Baltimore, MD 21218, USA
 	      	\and
             	University of Oxford, St John's College Research Centre, Oxford, OX1 3JP, UK
	  }

\date{Received  / Accepted  }

\authorrunning{E.~Costantini et al.}
\titlerunning{\xmm\ observation of \mrk}

\abstract{We present the results of a $\sim$160\,ks-long \xmm\ observation of the Seyfert~1 galaxy \object{Mrk~279}. 
The spectrum shows evidence of both broad and narrow emission features. The \fek\ line may be equally well explained by a
single broad Gaussian (FWHM $\sim10\,000$\,\kms) or by two components: an unresolved core plus a very broad profile
(FWHM$\sim14\,000$\,\kms). For the first time we quantified, via the ``locally
optimally emitting cloud" model, the contribution of the broad line region (BLR) to the absolute luminosity of 
the broad component of the \fek\ at 6.4\,keV. 
We find that the contribution of the BLR is only $\sim$3\%. 
In the two-line component scenario, we also evaluated 
the contribution of the highly ionized gas component, which produces the \fexxvi\ line in the iron K region. This 
contribution to the narrow core of the \fek\ line is marginal $<$0.1\%. Most of the
luminosity of the unresolved, component of \fek\ may come from the obscuring torus, 
while the very-broad associated
component may come from the accretion disk. However, models of reflection by cold gas are difficult to test because
of the limited energy band. The \fexxvi\ line at 6.9\,keV is consistent to be produced 
in a high column density ($N_{\rm H}\sim10^{23}$\,cm$^{-2}$), extremely ionized (log$\xi\sim5.5-7$) gas. 
This gas may be a highly
ionized outer layer of the torus.}


\maketitle

\section{Introduction}

The X-ray spectrum of active galactic nuclei (AGN) bears the signature of different environments in the vicinity of the
supermassive black hole. In particular, the emission features detected in an X-ray spectrum
are believed to arise in different environments, with dramatically distinct physical characteristics of temperature
and density. Narrow emission line profiles (especially from C, N, O He-like and H-like ions),
 not significantly variable on a short time-scale, may form in the far
narrow-line region \citep[e.g.][]{pounds}. 
The broad symmetric line profiles, the 
more and more often detected in AGN
spectra \citep[][hereinafter C07]{kaastra02,ogle04,steen05,costantini07}, are consistent to be produced, at least in one
case, within 
the broad line region (C07), which is $\approx5.6L^{1/2}_{42}$ 
light days away from the central black hole \citep[$L$ is the luminosity of \civ\ in units of
$10^{42}$erg\,s$^{-1}$,][]{peterson93}. 
Emission from the accretion disk, located in the proximity of the black hole,
 has been extensively studied via the iron K\,$\alpha$ line at 6.4\,keV. For this
line, relativistic effects result in a skewed and asymmetric line profile \citep[e.g.][]{tanaka95,young,wilms}.
Recently, relativistic profiles have been reported also for other lines, in particular \oviii\
\citep{graziella01,kaastra02,ogle04,steen09}. 
However, a contribution to the line emission from other AGN regions (NLR, BLR and molecular
torus) add up to these relativistic profiles. 
A narrow and not variable component, probably originating far from the black hole, 
has been disentangled from the Fe\,K\,$\alpha$ line
\citep{y_pad04,elena05}. This component is predicted in the unification model \citep{anto_miller85} 
as produced in a high-column
density, cold region, about 1\,pc away from the black hole \citep[e.g.][]{kk87,k94} and then scattered into our line
of sight \citep[e.g.][]{ghisellini94}. An unresolved \fek\ line and the associated reflection seems indeed to be ubiquitous
in bright type~1 objects \citep{nandra07}.\\   
As the FWHM of the narrow core is compatible with the width of the UV/optical broad lines \citep{y_pad04}, 
a contribution to the broad component from the BLR is in principle possible. 
However, considering individual sources, no evident
correlation between the FWHM of neither the hydrogen H$\alpha$ or H$\beta$, 
supposed to be entirely formed in the BLR, has been found \citep{sulentic98,nandra06}. 
These results are based on optical and X-ray data not simultaneously collected. In
\object{NGC~7213},
using simultaneous observations, the H$\alpha$ and the resolved iron line shared the same value of the FWHM \citep{bianchi08}.

\mrk\ is a bright Seyfert~1 galaxy ($F\sim2.55\times10^{-11}$\,erg\,cm$^{-2}$\,s$^{-1}$, this study), extensively studied in X-rays (C07 and
references therein). C07, using \ch-LETGS data of this source,
 for the first time quantified the contribution of the BLR to the soft X-ray spectrum. 
 Indeed, thanks to the simultaneous observation of the broad lines in the UV band by HST-STIS and FUSE \citep{gabel05} 
 the ionization structure of the BLR has been determined using the ``locally optimally
 emitting clouds" model \citep[LOC,][]{baldwin95}. This allowed to infer the X-ray broad lines luminosities, that were
 then contrasted with the LETGS data. C07 found that the broad lines observed in the soft X-ray spectrum (where the
 \ovii\ triplet was the most prominent one) were 
 consistent to be entirely formed in the BLR. The average peak of production of the X-ray lines is 
 possibly $\sim10$ times closer in with respect to the UV lines, implying a larger Keplerian width for the X-ray lines.  
 Unfortunately, the limited LETGS energy band prevented C07 to study the iron K\,$\alpha$ region. 
 Here we present a
 detailed study of the X-ray spectrum of \mrk\ as observed by \xmm, focusing mainly 
 on the emission components and the 6.4--7\,keV region. 

The paper is organized as follows. Sect.~2 is devoted to the spectral analysis of the data. In Sect.~3 we model the emission
spectrum evaluating, for each component, the contribution to the \fek\ line. In Sect.~4 we discuss our results and in
Sect.~5 we present the conclusions. 
The cosmological parameters used are: 
H$_0$=70 \kms\,Mpc$^{-1}$, $\Omega_m$=0.3, and $\Omega_{\Lambda}$=0.7. The abundances were set to solar following
\citet{ag89} prescriptions. The redshift is 0.0305 \citep{scott04}. 
The quoted errors are 68\% confidence errors ($\Delta\chi^2=1$), unless otherwise stated. 

\section{The data analysis\label{par:data}}
\begin{figure}
\resizebox{\hsize}{!}{\includegraphics[angle=90]{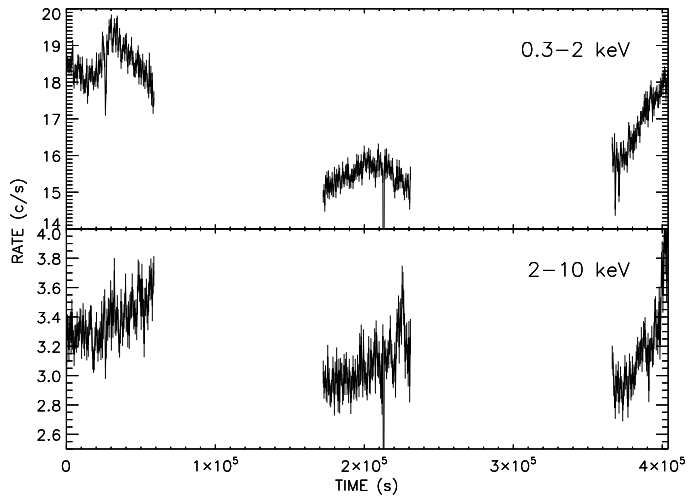}}
\caption{\epi\ light curves in the 0.3--2 and 2--10\,keV band. The maximal variation is about 36\%.}
\label{f:lc}
\end{figure}

The observation of \mrk\ was spread over 3 orbits on November 15--19 2005. In Table~\ref{t:obs_log} we report the observation log.
The data were processed with the standard
 SAS pipeline (SAS~7.0) and filtered for any background
flares. The \epi\ exposure time in small-window mode reduced the exposure time from the 
original 160\,ks to 110\,ks. After the background filtering we obtained a net exposure time of $\sim$75\,ks. 
The light curve of the three time intervals is displayed in Fig.~\ref{f:lc} in the soft (0.3--2\,keV) and hard (2--10\,keV)
band. The maximal variation is about 36\% in both energy bands.
To increase the statistics, we combined the \epi\ data after verifying that the physical parameters of the modeling were 
the same for the three separate data sets. Epic-MOS1 camera was set in timing mode. Due to still existing energy gain uncertainties in this mode, 
we did not analyze these data further. Epic-MOS2 were collected in imaging mode that resulted in about 61\,ks of net
observation, after filtering out short high background episodes. For both \epi\ and MOS2 we imposed at least 20 counts per channel to allow the
use of the $\chi^2$ minimization.\\ 
The RGS total useful exposure time is about 110\,ks. Each RGS data set has been rebinned by a factor 5 which resulted 
in a bin size of $\sim$0.07\AA\ and signal-to-noise ratio of 10 for each set of data that were simultaneously fitted. 
The spectral analysis was carried out using the fitting package 
SPEX\footnote{http://www.sron.nl/divisions/hea/spex/version2.0/release/} (ver~2.0).   

In the following we describe first the underlying continuum, then the emission line spectrum and finally 
the absorption features which furrow
the spectrum.\\

\begin{table}
\caption{\label{t:obs_log}\xmm\ observation log.}
\begin{center}
\begin{tabular}{llllc}
\hline
\hline
Orbit & Exp  & Exp net & date			& \epi\ rate\\
	& (ks) &  (ks) & dd-mm-yy &		c/s (0.3--10\,keV)\\
1087 &	59.3	& 27.4 &	$15-11-05$	& 25.7\\
1088 &	59.7 	& 32.0 &	$17-11-05$	& 21.6\\
1089 &  38.0  	& 15.5 &	$19-11-05$	& 23.5 \\
\hline
\end{tabular}
\end{center}
Note: The orbit number, the nominal exposure, the net exposure, the date and the 2--10\,keV \epi\
background-subtracted count rates are listed.
\end{table}

\subsection{The continuum}\label{par:cont}
\begin{figure}
\resizebox{\hsize}{!}{\includegraphics[angle=90]{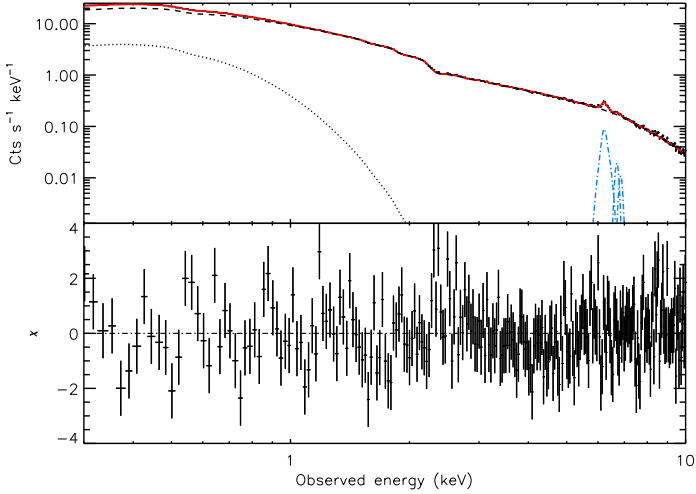}}
\caption{Best fit for the \epi\ data. The model includes a modified black body (dotted line), a broken powerlaw (dashed
line), both absorbed by ionized gas, a
broad \fek\ (see Sect.~\ref{par:broad}) line and two narrow lines at 6.9 and 7.05\,keV (dash-dotted lines).}
\label{f:pn_cont}
\end{figure}

The continuum has been first determined using \epi. A single powerlaw fit, modified by Galactic absorption is
unacceptable ($\chi^2/\nu>34$, where $\nu$ is the number of degrees of freedom). This fit produces 
systematic residuals both at soft and hard energies, in addition to narrow features both in absorption and emission. 
We then added
a black body component, modified by coherent Compton scattering
\citep{kaastrabarr}. The reduced $\chi^2$ is again unacceptable ($\chi^2/\nu\sim6$). The residuals show that a broad band component is still missing. We then substituted the
simple powerlaw model with a broken power law, obtaining an acceptable description of the continuum. The reduced
$\chi^2$, considering only the best-fit continuum, without narrow emission/absorption features, is $\chi^2/\nu\sim2.5$. 
The parameters of this phenomenological interpretation of the 
continuum are listed in Table~\ref{t:continuum} and the fit shown in Fig.~\ref{f:pn_cont}. 
The parameters in Table~\ref{t:continuum} and reduced $\chi^2$ refer to the total best fit, including emission and
absorption features, as described below (Sect.~\ref{par:broad}, \ref{par:abs}).\\ 
For comparison, we analyzed also MOS2 data, with the caveat that deviation from the \epi\ best fit may occur,
especially at the low (E$<$0.5\,keV) and high energy (E$>$8\,keV) ends of the band, and at the instrumental gold edge
around 2\,keV\footnote{Several examples are provided at http://xmm2.esac.esa.int/cgi-bin/ept/preview.pl}. The final MOS2 best-fit parameters are 
shown in Table~\ref{t:continuum}. The parameters agree well with the \epi\ fit, albeit with some differences, 
which we ascribe to the cross calibration mismatch between the MOS and PN cameras.
\begin{table}
\caption{\label{t:continuum}Best-fit continuum parameters for a phenomenological model. }
\begin{center}
\begin{tabular}{llll}
\hline\hline

&& PN & MOS2\\
\hline
$T_{\rm mbb}$ & keV & $0.16\pm0.02$ & $0.12\pm0.02$\\
$\Gamma_1$ & &$2.08\pm0.01$ & $2.03\pm0.02$\\
$\Gamma_2$ & &$1.75\pm0.01$ & $1.82\pm0.03$\\
$E_0$ & keV& $2.31\pm0.04$ & $2.0\pm0.1$\\
$\chi^2/\nu$ && 281/234 & 262/208\\
\hline
\end{tabular}
\end{center}
Note: \epi\ and MOS2 fitted separately. The reduced $\chi^2$ refer here to the final best fit.
\end{table}
\subsection{The emission line spectrum}\label{par:broad}
\begin{figure}
\resizebox{\hsize}{!}{\includegraphics[angle=90]{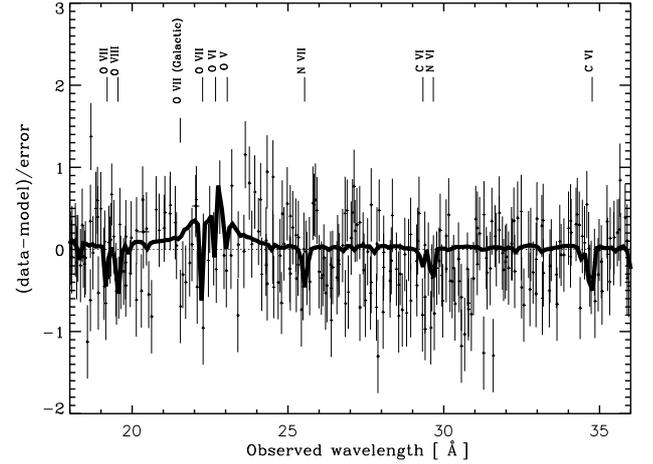}}
\caption{\label{f:rgs} RGS spectrum in the \ovii\ region in terms of ratio from a continuum fit.
For display purpose, only the rebinned data from orbit 1087 and 1088 are shown. 
The measurements reported in the paper are based on the full RGS exposure. 
The labels indicate some of the main features of the warm absorber.}
\end{figure}

\begin{figure}
\resizebox{\hsize}{!}{\includegraphics[angle=90]{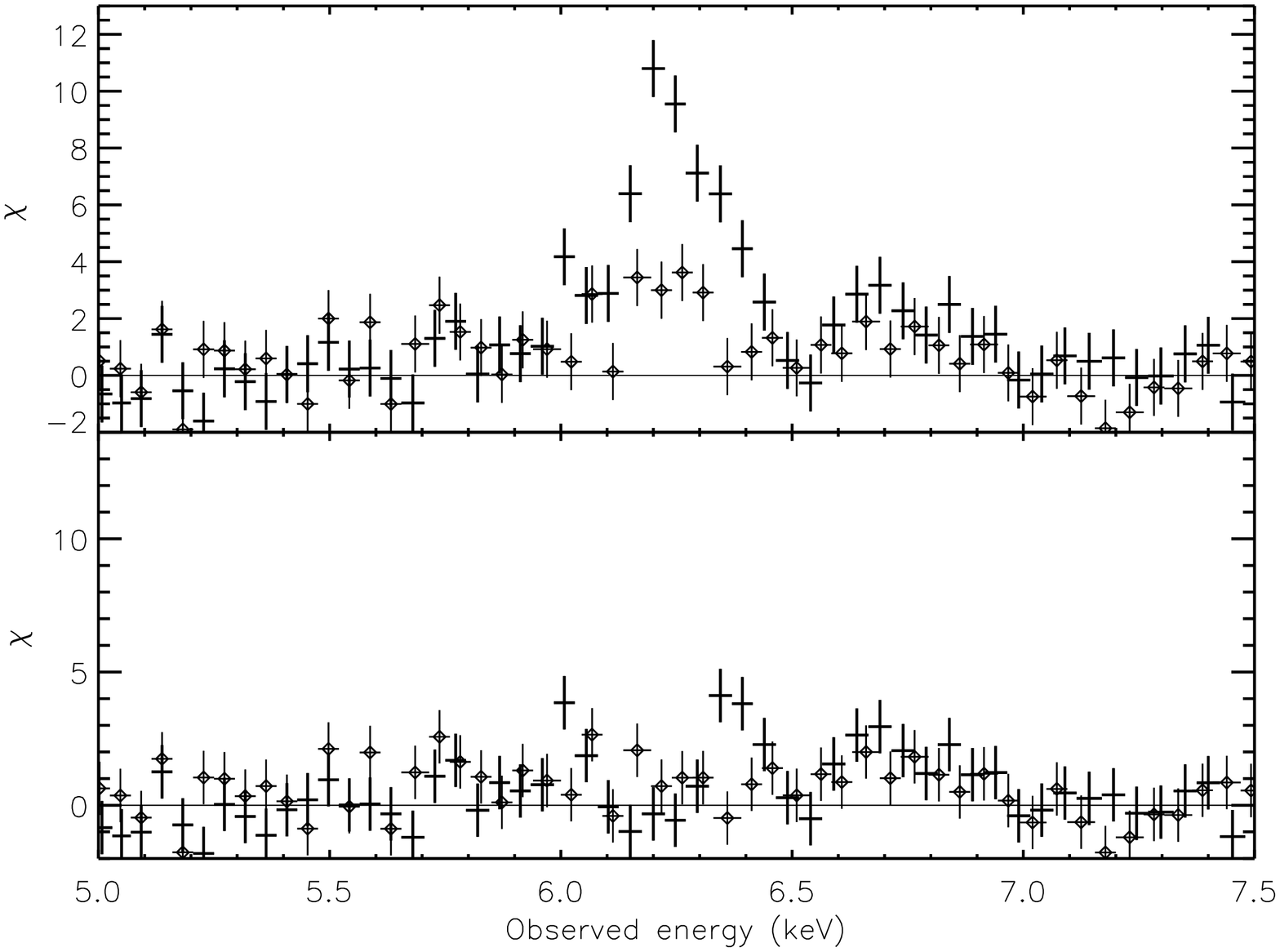}}
\caption{\label{f:fek}Upper panel: residuals, in terms of $\sigma$ to the continuum in the iron line region as observed by \epi. 
Three features are evident: a prominent \fek\ 
line and weaker lines identified as \fexxvi\ and 
Fe\,K$\beta$. Lower panel: residuals, in terms of $\sigma$ after the inclusion of a narrow line at 6.4\,keV. The \epi\ data
are identified by simple crosses, and the MOS2 data by diamonds.}
\resizebox{\hsize}{!}{\includegraphics[angle=90]{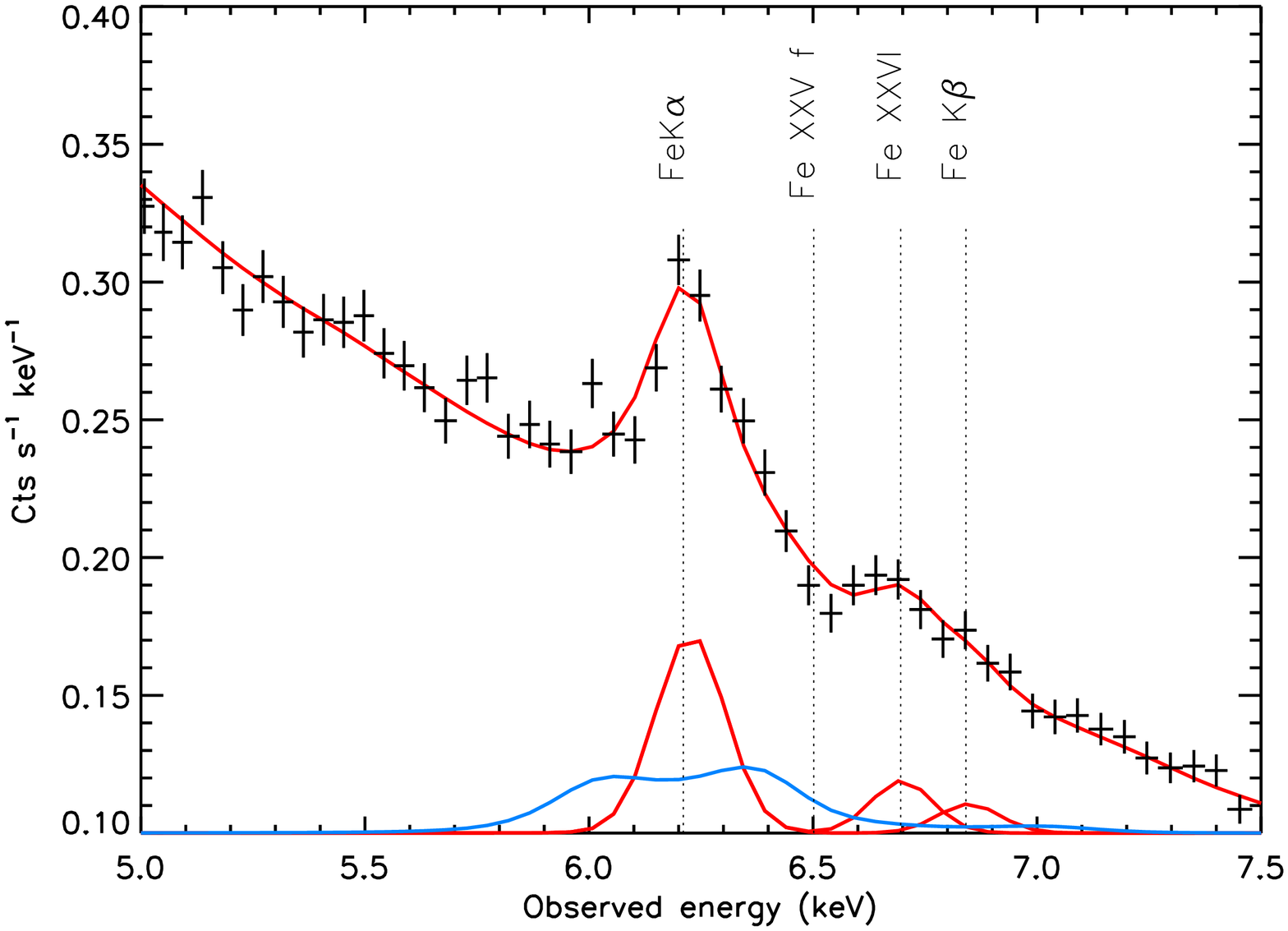}}
\caption{\label{f:fit_fek}Spectrum and best fit in the iron line region as observed by \epi. 
The solid lines display the broad and narrow
components added to fit the emission features: a prominent \fek\ 
line and weaker lines identified as \fexxvi\ and 
Fe\,K$\beta$.}
\end{figure}

We see evidence of a range of broad and narrow emission lines both in the RGS and in the Epic spectra. 
Here we refer as narrow
lines the ones which are unresolved by \epi\ (i.e. FWHM$<$7000\,\kms), but may be resolved by RGS. 
Then the broad lines (FWHM$>$7000\,\kms) are instead resolved also by Epic. 
Finally, the very broad lines are the ones with 
FWHM$\sim$14\,000\,\kms. The only example of the latter in the present data is the broad, 
possibly relativistically smeared, component of the iron line at
6.4\,keV.\\
In our previous study broad emission from a blend of the \ovii\ triplet lines was very significantly detected 
(at $\sim6\sigma$, C07). 
We therefore looked for such evidence in the present RGS data. We fix the wavelength of the line 
centroid ($\lambda$=21.9\,\AA) to the values found in C07. We also fix the 
FWHM of the broad feature, which could not be resolved in the individual \ovii\ lines, 
to the previously measured value: 1.4\,\AA, corresponding to about 19\,000\,\kms\ for the blend. 
We find that the intrinsic line luminosity decreased with respect to the 2003 
observation going from $50\pm8$ to $13\pm7$ in units of
$10^{40}$\,erg\,s$^{-1}$. The same line luminosity is measured also using the three RGS data sets separately. 
The inclusion of the line improves the fit by $\Delta\chi^2/\Delta\nu\sim6/1$, corresponding to only $2.4\sigma$ significance. 
The broad line
is also a necessary emission component to correctly fit some underlying absorption features (e.g. \ov, \ovi\ and \ovii), 
important in the global fitting of the
ionized absorber parameters (Fig.~\ref{f:rgs}). The inclusion of this low-ionization absorber alone 
improves the fit of the RGS spectrum by $\Delta\chi^2/\Delta\nu\sim58/3$.\\  
In the RGS band we detect the narrow \ovii\ forbidden line with a flux
($9\pm6\times10^{-15}$\,erg\,cm$^{-2}$\,s$^{-1}$), consistent with the \ch\ measurement (C07). The line is marginally 
detected (Table~\ref{t:narrow}), probably because of a neighboring bad pixel. 
For other relevant narrow lines, such as 
\oviii\ \Lya, \neix\ and \nex, we only obtain upper limits on the luminosity (Table~\ref{t:narrow}).\\ 
Both the PN and MOS2 spectra show a complex structure between 6.4 and 7\,keV: 
a prominent Fe\,K\,$\alpha$ line at $\sim$6.4\,keV (rest-frame) and a broad emission feature, consistent with a blend of 
\fexxvi\ and \fekb\ at $6.90\pm0.06$\,keV and $7.05\pm0.05$\,keV respectively (Fig.~\ref{f:fek}, upper panel). 
Because the MOS2 spectral analysis deliver a slightly different powerlaw slope at energies $>2$\,keV
(Table.~\ref{t:continuum}), and the data are affected by lower statistics, we restrict
the iron line analysis to the \epi\ data only. At first we modeled the \fek\ line with a single unresolved Gaussian. Although the general fit improves 
($\chi^2/\nu=371/243\sim1.5$) This model does not provide a good fit for the line, 
as evident in Fig.~\ref{f:fek} (lower panel).  
Next, we considered the line as composed of an unresolved Gaussian and a very broad component, 
leading to a significant improvement of the fit ($\Delta\chi^2/\Delta\nu=32/3$, Table~\ref{t:fek}). 
The very broad component has a FWHM of $0.29\pm0.04$\,keV, corresponding to $\sim$14\,000\,\kms. 
We tested this very broad component of the \fek\ line 
against a relativistically smeared profile \citep{laor}. With this model, the line 
should be viewed at an angle of $\sim15^{\circ}$, 
with a radial emissivity law $r^{-q}$, with $q\sim1.5$ (Fig.~\ref{f:fit_fek}, Table~\ref{t:fek}). 
These parameters mimic a regular, almost symmetric, line
profile \citep[e.g.][]{reynolds03}. The statistical improvement with respect to a simple very-broad Gaussian is: 
$\Delta\chi^2/\Delta\nu\sim5/1$, corresponding to $\sim2.2\sigma$.  
For the \fek\ feature, we considered also a single Gaussian with the width free to vary (Table~\ref{t:narrow}). The line centroid is $6.41\pm0.08$ in the rest
frame of the source, and the
intrinsic luminosity is $61\pm6\times10^{40}$\,erg\,s$^{-1}$. The line is resolved by \epi, with a FWHM of
$0.21\pm0.03$\,keV (corresponding to $\sim10\,000\pm 1000$\,\kms). The statistical improvement with respect to a single
unresolved line is $\Delta\chi^2/\Delta\nu=37/1$.\\
The inclusion of other two narrow lines, at the energy of the \fexxvi\ and the \fekb\ further improves the fit by  
$\Delta\chi^2/\Delta\nu=31/2$. 
For consistency, we tested the Laor model, with the same line parameters, 
to the \fekb, fixing the line ratio to 0.135 \citep{yaqoob07, palmeri}. 
As expected the flux of a broad \fekb\ line has a negligible effect on the fit, therefore we ignore this additional
component in subsequent fits. This negligible contribution of a broad profile of the \fekb\ also strengthens our assumption that the bump seen in the 6.9--7.05 region is a blend of
two narrow lines (\fexxvi\ and \fekb) rather than a broad \fekb\ line arising from the accretion disk. 
This fit is shown in Fig.~\ref{f:fit_fek}.\\
The reduced $\chi^2$ of a fit with all emission lines included, but no absorption, is
308/238. With this final component (see below), the $\chi^2/\nu=281/234\sim1.2$. The lines' parameters and their significance are listed in Table~\ref{t:narrow}. In all the lines fitting, 
we also let the normalization free to vary towards negative values. This allowed us to use the F-test as an indication of the
significance of the lines \citep{protassov}.    
\begin{table}
\caption{\label{t:narrow}list of line luminosities as measured by Epic-PN (inst.~1) and RGS (inst.~2). 
}
\begin{center}
\begin{tabular}{lllll}
\hline
\hline
ion	&	Rest & Lum$_{\rm obs}$	& F-test & inst. \\
	& wavelength &				& &\\
	&	(\AA)		& ($10^{40}$ erg s$^{-1}$)&\%&	\\
\hline	
{\bf narrow lines}	&	&&\\
\hline
Fe\,K$\beta$	& 1.75		& $6.6\pm2.8$&89.0&1\\
\fexxvi\,\Lya	& 1.77		& $10.0\pm2.7$&99.98&1\\
\fexxv\ f.	& 1.85		& $<4.7$&$\dots$&1\\
\fek$^1$	& 1.93		& $12.0\pm7.0$&$>$99.99&1\\
\nex\ 		&12.14		& $<1.2$&$\dots$&2	\\
\neix\ f.	& 13.69		& $<1.25$&$\dots$&2	\\
\oviii\,\Lya	&18.97		& $<1.0$&$\dots$&2\\
\ovii\ f.	&22.10		& $3.1\pm2.0$&89.0&2	\\
\nvii\,\Lya	&24.78		& $<3.1$&$\dots$&2\\
\nvi\ f.	&29.53		& $<1.18$&$\dots$&2\\
\hline
{\bf broad lines} &	&&\\
\hline
\ovii\ triplet$^2$ & 21.9 & $13\pm7$ &95.0& 2\\
\fek$^3$ & $1.93\pm0.03$ & $61\pm6$ & $>$99.99 & 1\\
\hline
\end{tabular}
\end{center}
Notes: We list the ion, the rest-frame wavelength, the
intrinsic luminosity and their significance, for narrow and broad lines.\\$^1$ Only narrow, unresolved component in a two-lines model.\\$^2$ Blend of the \ovii\ triplet lines.\\$^3$ Single Gaussian with free
line-width model.  
\vspace{0.5cm}
\renewcommand{\tabcolsep}{1mm}
\caption{\label{t:fek} Parameters for the very broad component of the \fek\ line in a two-components model. 
}
\begin{center}
\begin{tabular}{llll}
\hline
\hline
 && Gauss$^1$ & Laor$^2$\\
\hline
$E$ &keV & $6.41\pm0.01$ & $6.47\pm0.03$\\
FWHM &eV & $0.29\pm0.04$ & $\dots$\\
Lum &$10^{41}$\,erg\,s$^{-1}$ & $4.64\pm0.16$ & $5.0\pm0.7$\\
$q$ & &$\dots$ & $1.5\pm0.5$\\
$i$ &deg & $\dots$ & $15\pm7$\\
$\chi^2/\nu$ && 339/240 & 334/239\\
\hline
\end{tabular}
\end{center}
Notes: Third column: one Gaussian with free width. 
Fourth column: Profile including relativistic effects. 
The reduced $\chi^2$ refers to an intermediate fit (see Sect.~\ref{par:broad})\\
$^1$ We list the energy of the centroid, the luminosity of the line and the FWHM.\\
$^2$ We list the energy of the centroid, the luminosity of the line, the emissivity law, 
$q$ and the inclination angle, $i$.
\end{table}
\begin{table}
\caption{\label{t:abs} Best fit parameters for the absorbed spectrum as measured by RGS and \epi.}
\begin{center}
\begin{tabular}{llll}
\hline
\hline
comp. & Param. & RGS & PN \\
\hline
1 	& 	$N_{\rm H}$ 	&	$0.7\pm0.2$ & $1.5\pm0.5$ \\		
	&	log$\xi$	&	$0.8\pm0.3$ &$1.1\pm0.3$ \\
	&	$v_{\rm out}$	&	$-300\pm150$ & --300 fix. \\
2	&	$N_{\rm H}$	&	$3\pm1$ & $4\pm1$ \\	
	&	log$\xi$	&	$2.6\pm0.1$ & $1.8\pm0.1$ \\
	&	$v_{\rm out}$	&	$-400\pm150$ & --400 fix. \\
\hline
\end{tabular}
\end{center}
Note: For each component we list the 
column densities ($N_{\rm H}$) in units of $10^{20}$\,cm$^{-2}$, ionization parameter (log$\xi$) 
and outflow velocities ($v_{\rm out}$, in km\,s$^{-1}$).
\end{table}
\subsection{The underlying absorption}\label{par:abs}

\mrk\ shows a complex absorption spectrum (C07), whose characteristics are best studied using RGS. 
In analogy  with C07, we modeled the absorption with two Galactic components,
that is a neutral gas \citep[$N_{\rm H}=1.64\times10^{20}$,][]{elvis89} and a ionized gas, 
highlighted by the \ovii\ absorption line at 21.6\AA, with $N_{\rm
H}\sim3.6\times10^{19}$\,cm$^{-2}$ and temperature $T\sim7.2$\,eV \citep[C07,][]{will05}. 
Further, we detect other two, photoionized, gas components intrinsic to the source. 
We report the average physical parameters of the warm absorbers as measured by RGS and the \epi\ in Table~\ref{t:abs}. 
The absorbers are characterized by a low $N_{\rm H}$. Therefore absorption lines,
 well detected in the high-resolution
spectrum (Fig.~\ref{f:rgs}), rather than edges 
(which are in this case shallow features in the Epic spectrum), are the signature of the absorbers. 
The inclusion of a lower ionization gas alone already improves the \epi\ fit by $\Delta\chi^2/\Delta\nu=17/2$.
The inclusion of a two-component gas, with column densities and ionization parameters free to vary leads to an 
improvement of the fit of $\Delta\chi^2/\Delta\nu=27/4$, bringing the reduced $\chi^2/\nu$ to $\sim1.20$. 
A detailed analysis of the complex absorption and a comparison with previous 
results will be presented in a forthcoming paper (Ebrero et al., in prep.). 
\begin{figure}
\resizebox{\hsize}{!}{\includegraphics[angle=90]{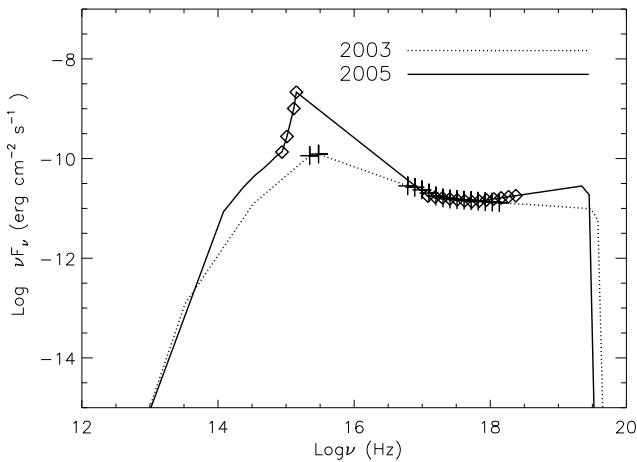}}
\caption{The SED of \mrk\ in 2005 measured by \xmm\ (solid line). The optical data are measured by OM, while the broad band 
X-ray continuum is measured by Epic-PN. The dotted line shows the SED of the 2003 data for comparison (C07).}
\label{f:sed}
\end{figure}
\section{Modeling of the \fek\ line region}

In this section we evaluate the contribution of different region in the proximity of the AGN 
to the spectrum above 6\,keV.

\subsection{The LOC model for the broad emission lines}\label{par:loc}
Here we investigate a possible physical link between the \fek\ line 
and the optical BLR. 
In order to quantitatively evaluate the contribution of the BLR to any X-ray broad lines, we used the ``locally optimally
emitting clouds" model \citep{baldwin95}. This model considers the observed line luminosity 
as a sum of lines emitted at different 
density $n$ and distances $r$, weighted by a powerlaw distribution for $n$ and $r$: $n^{-\beta}$ and $r^{-\gamma}$
respectively \citep[see e.g. C07,][for details]{kor97}. 
For the present observation we can benefit from the results obtained in C07: 
they evaluated the ``structure" of the BLR (i.e.
the distribution of $r$ and the covering factor) fitting 11 UV lines observed by HST-STIS and FUSE simultaneously to the
\ch-LETGS data. They found a slope for $r$, $\gamma=-1.02\pm0.14$ and a covering factor $C_V=34\pm26$\%, keeping the slope for $n$
fixed to --1 \citep{korista00}. 
The general structure of the BLR is not expected to have changed significantly in the 2.5 years that have
elapsed since the \ch\ observation. Therefore we apply those value also to the present spectral energy distribution (SED).

However, the ionizing continuum, and as a consequence the line luminosities, may have changed. In Fig.~\ref{f:sed} the
SED used for the present observation (2005) is displayed (solid line). The optical points are measured
by OM using the U (344\,nm), UVW1 (291\,nm), UVM2 (231\,nm) and UVW2 (212\,nm) filters. 
The unabsorbed X-ray continuum is evaluated using the Epic-PN data.  
For energies higher than 10\,keV, the power law continuum was extended and artificially cut off at
$\sim$150\,keV. On the very low energy end of the SED (far infrared to radio band), 
the shape was taken from a standard AGN SED
template used in Cloudy \citep{fer98}. We used Cloudy (version~07.02.02) to obtain a line luminosity grid over
 a large range of $r$ and $n$
values. As in C07, $r$ ranged between $10^{14.7-18}$\,cm. The value of $n$ ranged between $10^{8-12.5}$\,cm$^{-3}$. 
For both parameters the spacing of the values was 0.125 in log.
For comparison, we show also the SED used for the 2003 observation. 
The overall optical flux is higher for the present data, while 
the X-ray continuum remained almost unchanged in flux and soft X-ray spectrum.\\

\begin{figure}
\resizebox{\hsize}{!}{\includegraphics[angle=90]{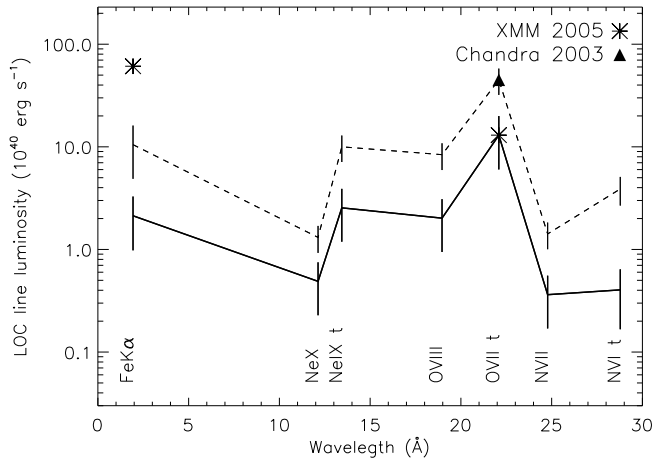}}
\caption{The solid thick line shows the predicted BLR line luminosities in the X-ray band in units of $10^{40}$\,erg\,s$^{-1}$ using the LOC model, using 
as constraints $C_V=34\pm26$\% and the observed luminosity of the broad \ovii\ line. The value of $\gamma$ is consistent
with what previously found for \mrk. The asterisks are the value of the observed broad line luminosities by XMM 
(Table~\ref{t:narrow}). The dashed line shows the LOC modeling based on the 2003 SED and observed \ovii\ line luminosity. 
The filled triangle shows the \ovii\ broad line luminosity observed by \ch.}
\label{f:loc_gamma}
\end{figure}

In order to infer the BLR contribution to the \fek\ line, we have to rely on the \ovii\ triplet in the X-ray band, whose 
luminosity could be totally accounted for by the LOC model (C07). 
Therefore we assume that all the emission from the 
broad line of the \ovii\ triplet comes from the BLR. 
The \ovii\ line luminosity 
($L_{\rm OVII}=13\pm7\times10^{40}$\,erg\,cm$^{-2}$\,s$^{-1}$) from the RGS data can be explained by the LOC model using
the new 2005 SED, with the values of $\gamma$
and $C_V$ set by the previous analysis, within the errors. Given these constraints, the
resulting value for $\gamma=1.17\pm0.03$, which is here the only free parameter, is found to be consistent with previous results.  
Once this constraint is set, we can
predict also the contribution of the \fek\ and other X-ray lines.\\ 
In Sect.~\ref{par:broad} we have provided two statistically acceptable descriptions of the \fek\ line: a
narrow$+$very-broad component model and a single, resolved, Gaussian. Only for the latter case the FWHM is similar to the
one directly measured from the UV data \citep[FWHM$_{\rm UV}$=8500--9500\,\kms,][]{gabel05}. 
In Fig.~\ref{f:loc_gamma} we show the BLR line
luminosities predicted by the LOC model in the X-ray band. Along with the \ovii\ observed 
luminosity we also plot the observed value for the \fek\ 
line in the case of a single Gaussian model (Table~\ref{t:narrow}). 
The predicted \ovii/\fek\ luminosity ratio is about 6.3 within the narrow 1.14-1.20 range for $\gamma$. 
Therefore, the BLR contribution to the \fek\ line is then 
$L_{{\rm BLR}}({\rm Fe\,K\alpha})=2\pm1\times10^{40}$\,erg\,s$^{-1}$, 
about 30 times smaller than the broad \fek\ component measured from the data. For comparison we also made a
prediction on the iron line flux based on the 2003 SED and the BLR parameters derived in C07. In that case the \fek\
luminosity of iron from the BLR is $10\pm5\times10^{40}$\,erg\,s$^{-1}$.

\subsection{Variability of the continuum and iron line complex}\label{par:var}
Lines emitted at a few gravitational radii from the black hole may show
significant short term variability. We examined the 
\fek\ line parameters in the three time segments of our observation using a simple Gaussian model. 
We find that the flux, the FWHM, centroid energy and the underlying
powerlaw slope are nearly the same in response to the central-source variation. 
In order to further test a variability of the \fek\ flux we analyzed both an archival \xmm-pn 
data set with a net exposure time 
of $\sim$26\,ks, collected in May 2002 and \ch\ HETGS data, collected about 
10 days later in May 2002 \citep{scott04,y_pad04}. 
In Table~\ref{t:fe_var} we show the comparison of the line parameters of
all observations since 2002, for a single-line model.\\
Prior to 1994, \mrk\ showed a significant 2--10\,keV flux variation, 
ranging from $1-5\times10^{-11}$\,erg\,cm$^{-2}$\,s$^{-1}$ 
\citep{weaver01}. The 2002 \ch-HETGS observation, revealed a very low-flux state of the source
\citep[$\sim1.2\times10^{-11}$\,erg\,cm$^{-2}$\,s$^{-1}$,][]{scott04}, 
a factor of two lower than what measured by \xmm, shortly before. 
In the last \xmm\ observation the source shows a 2--10\,keV flux of
$\sim2.5\times10^{-11}$\,erg\,cm$^{-2}$\,s$^{-1}$. 
 We see that, 
despite the change in the continuum flux (a factor two), the \fek\ line 
parameters did not dramatically change in three years time, within the errors.\\
In the \ch\ data the decline of the HETGS effective area and the low signal-to-noise ratio hampered 
the detection of the \fexxvi\ and Fe K\,$\beta$ lines. On the contrary, in the 2002 \xmm\ spectrum, both lines, although blended, are
detected at about 95 and 90\% confidence level for \fexxvi\ and Fe K\,$\beta$, respectively. 
The line luminosities are the same we obtain in 2005, within the errors. 
\begin{table}
\caption{\label{t:fe_var} Parameters of the \fek\ at different epochs.}
\begin{center}
\begin{tabular}{llll}
\hline
\hline
year & $E$ & $L$ & FWHM\\
& (keV) & (10$^{40} $erg\,s$^{-1}$) & (keV)\\
\hline
2002a & $6.43\pm0.03$ & $46\pm11$ & $0.26\pm0.12$\\
2002b & $6.42\pm0.04$ & $41\pm15$ & $<0.20$\\ 
2005 obs 1 & $6.42\pm0.01$ & $56\pm8$ & $0.19\pm0.05$\\ 
2005 obs 2 &$6.40\pm0.02$ & $53\pm8$ & $0.17\pm0.05$\\
2005 obs 3 & $6.42\pm0.03$ & $86\pm17$ & $0.3\pm0.1$\\
\hline
\end{tabular}

\end{center}
Note: We list the energy, intrinsic luminosity and FWHM 
for a single-line fit to the \fek\ profile for the spectra obtained in
2005 in three separate orbits and for observations collected in 2002 by \xmm\ (2002a) and \ch-HETGS (2002b).
\end{table}
    
\begin{figure}
\resizebox{\hsize}{!}{\includegraphics[angle=90]{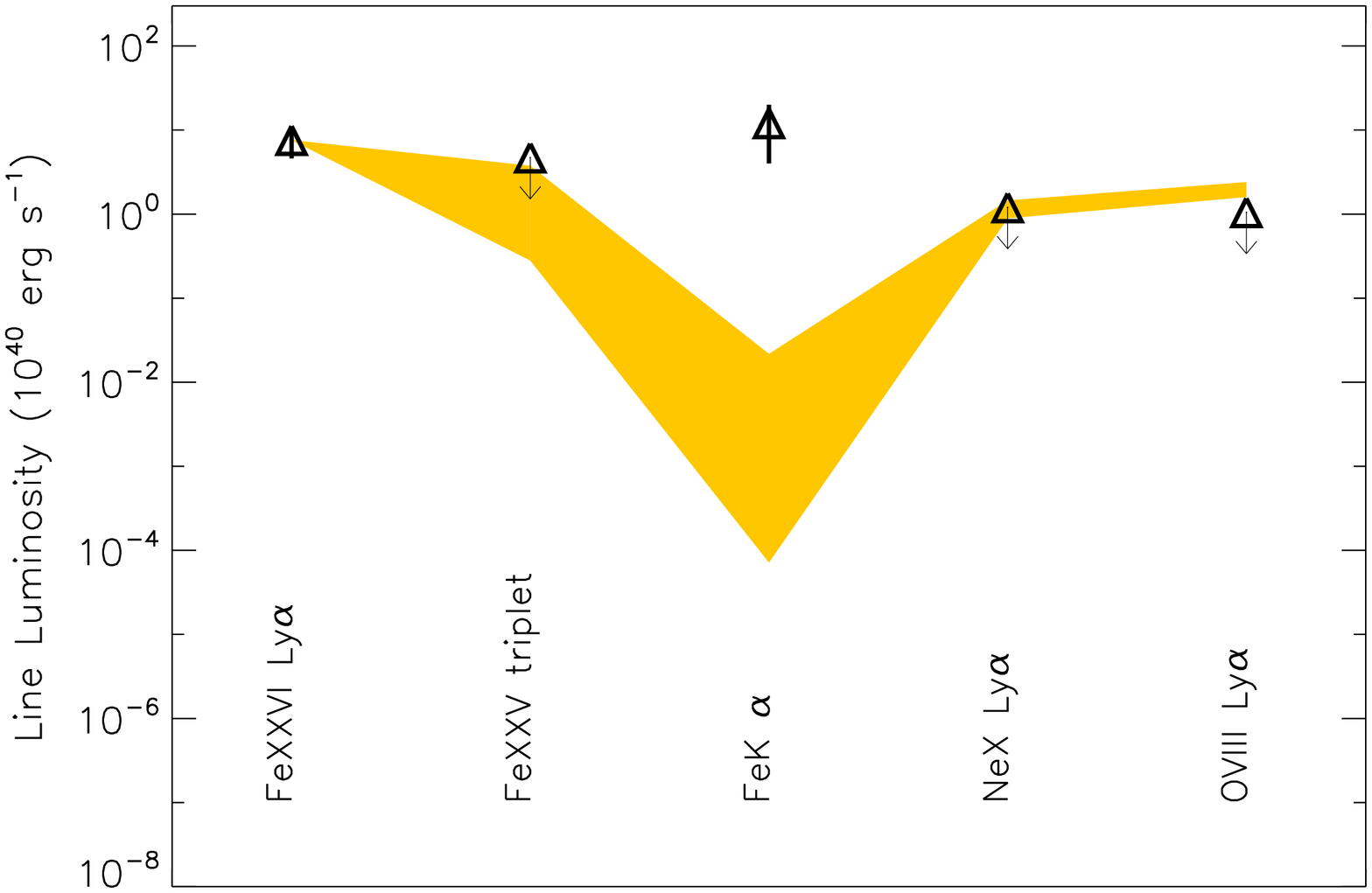}}
\caption{\label{f:high}Emission line contribution to the high ionization lines (\fexxvi, \fexxv, \nex, \oviii). 
Triangles: data. Light
shaded line: range of models viable to fit the data.}
\resizebox{\hsize}{!}{\includegraphics[angle=90]{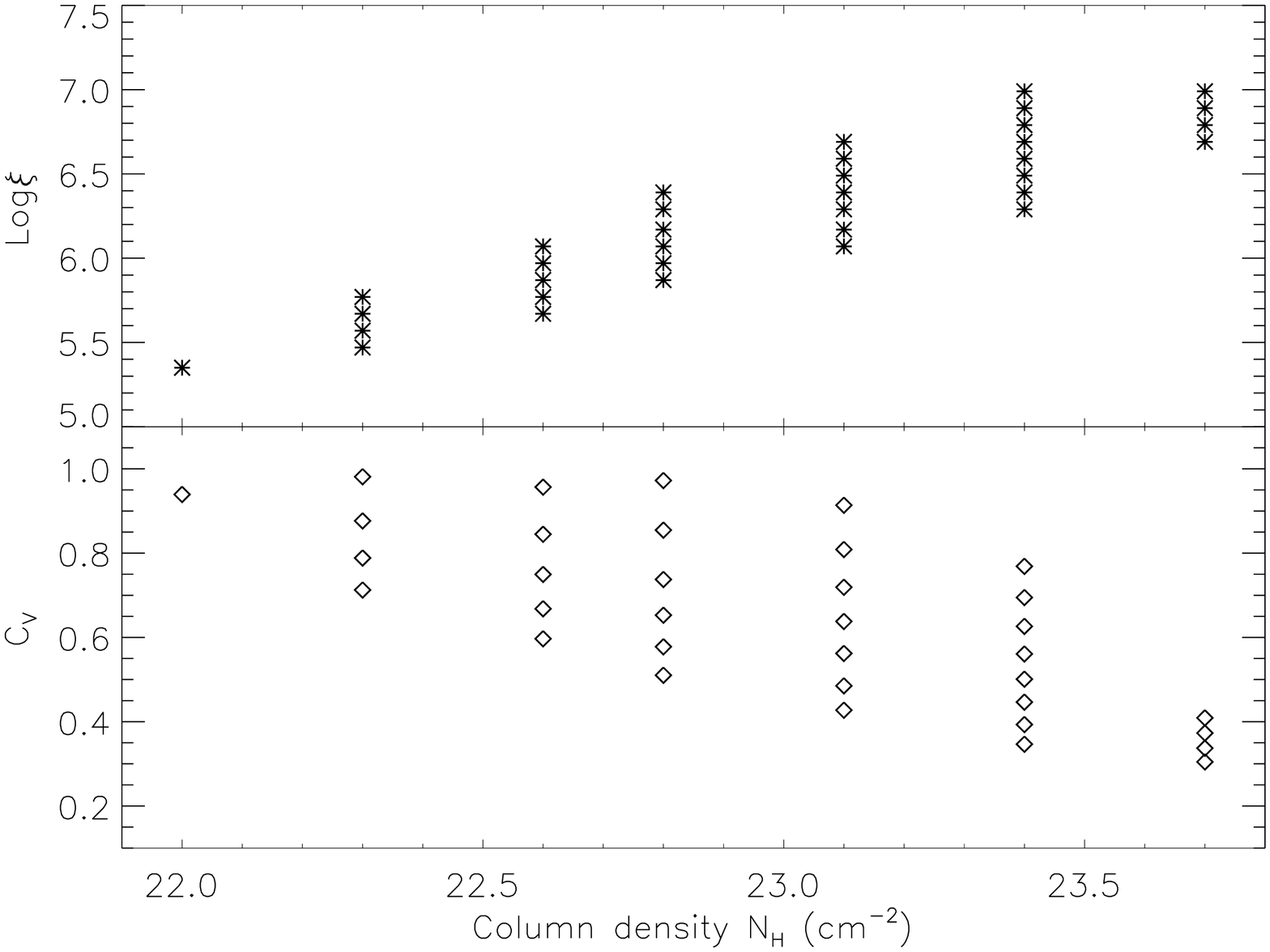}}
\caption{\label{f:high_xi_cv}Parameter space of parameters viable to produce 
the observed high-ionization emission spectrum. Upper panel: Log of the ionization parameter as a function of 
the log of the column
density. Lower panel: Covering factor of the \fexxvi\ line as a function of the log of the column density.}

\end{figure}
\subsection{The \fexxvi\ line at 6.9\,keV}\label{par:high}

There are no detectable changes in the \fexxvi\ line in a 3-years time scale. The absence of any of the 
lines (or a blend of lines) of the \fexxv\ triplet is also peculiar (Table~\ref{t:fek}, Fig.~\ref{f:fit_fek}). 
In order to understand the properties of the gas emitting \fexxvi, we created a grid of
column density, log$N_{\rm H}$= 22.0--24.5, and ionization parameter, log$\xi$=3--7, using Cloudy. 
For lower-column-density, lower-ionization
parameter gas, the predicted emission line would be too weak, or absent. 
For an easy scaling of the luminosity, 
we initially considered the covering factor ($C_V$) equal to unity. This is the fraction of light that 
is occulted by the absorber\footnote{This is different from the 
global covering factor, the fraction
of emission intercepted by the absorber averaged over all lines of
sight, which is has been estimated to be about 0.5 \citep[e.g.][]{crenshaw2003}. }. A covering factor of one 
is of course unrealistic, 
because in this case absorption lines of the same ions should be observed. Therefore $C_V$ must be less than one. 
We also simplistically assume that a single gas component is responsible for all \fexxvi\ emission. 
Therefore, we took the measured intrinsic luminosity of \fexxvi\ as the reference line to estimate the covering factor.
We only have upper limits on other highly ionized lines (namely \nvii, \oviii, \nex\ and \fexxv, Table~\ref{t:fek}), 
therefore a formal $\chi^2$ fit cannot be performed.
However, those limits contain information that can be used to constraint the parameters. In Fig.~\ref{f:high} 
the set of parameters which are consistent with the measurements are displayed.\\ 
In Fig.~\ref{f:high_xi_cv}, we show the parameter space of the possible solutions.
We see that the ionization parameter sufficient to produce
\fexxvi, but not \fexxv\ has to be larger that $\sim$5.3. The ionization parameter linearly 
correlates with the column density, 
as for the highest values of log$\xi$ the emission lines are visible only if there is sufficient emitting material. 
For the same reason, for the same column density, log$\xi$ and $C_V$ correlate linearly.    
On the other hand, the covering factor anti-correlates with $N_{\rm H}$ (Fig.~\ref{f:high_xi_cv}). 
Indeed, for a high column density less material on the line of sight is needed to produce the observed emission.
\section{Discussion}\label{par:dis}

\subsection{A complex continuum}\label{par:reflection}
The continuum spectrum of \mrk\ needs at least three component to be correctly interpreted: a modified black body and a
broken powerlaw (Table~\ref{t:continuum}). In principle, reflection from both the accretion disk and from
distant matter should be present \citep{nandra07,k_kriss95,matt_perola_piro91,george91}. 
The only difference would be that if arising from the accretion disk, 
the reflection and the associated iron line should be modified by relativistic effects. 
The harder tail with $\Gamma\sim1.75$ and the double component of the iron line profile (Sect.~\ref{par:broad}) 
may be reminiscent of a reflection emission component.  
We therefore tested this scenario, in comparison with our phenomenological model for the continuum, including both types of reflection
in the model (model REFL in SPEX). This model simultaneously considers the Compton reflected continuum \citep{mz95}
and the fluorescent emission from \fek\ \citep{zc94,z99} from a Schwarzschild black hole. 
General relativity effects and convolution with an accretion
disk effects can be easily switched off in this model. 
Free parameters are the normalization of the reflected power law and
its spectral index $\Gamma$, the reflection scale\footnote{The parameter $s$ is a scaling factor for the reflected
luminosity: $L_{\rm tot}=L_{\rm pow}+s L_{\rm refl}$. For an isotropic source above the disk $s=1$, corresponding to an equal
contribution from direct and reflected spectrum.} $s$, the emissivity scale $\alpha$ \citep{z99}, 
the inclination angle $i$ of the disk of the iron line emission. 
In the unblurred reflection, we fixed the inclination angle
to 60$^{\circ}$, based on the average found for a large sample of Seyferts \citep{nandra07}.\\
We find that the energy range $E>3$\,keV 
can be well fitted by a combination of a reflector with relativistic properties and by a distant reflector. 
The former accounts for the broad
component of the \fek\ line, while the latter models the narrow 
component of the line (Table~\ref{t:refl}) 
in a similar way as obtained in the two-component fit (Fig.~\ref{f:fit_fek}).\\
The basic parameters of the disk (e.g. the inclination angle, Table~\ref{t:refl}) are in agreement with what was found from the \fek\ line fit 
using the \citet{laor} model. We note
however that here a comparison is not straightforward as in \citet{zc94} a diskline
 from a Schwarzschild (rather than Kerr) 
black hole 
is included.\\ 
The two reflection component cannot fit simultaneously the broad band continuum, 
as the soft energy portion of the spectrum is significantly underestimated. 
To reach an acceptable fit, this region has to be modeled again 
by a modified black body and a single powerlaw (Table~\ref{t:refl}). 
Residuals mainly at the crossing points of the continuum components determine the value of the reduced $\chi^2$ 
(Table~\ref{t:refl}). Therefore a broad band modeling only in terms of reflection is not straightforward. In addition, 
hard X-rays coverage is not available and this fit is based on the 0.3--10\,keV
continuum shape and an iron line with two blended components. This limited knowledge on the broad band behavior 
makes the fit parameters very uncertain.  
\begin{table}
\caption{\label{t:refl} Alternative fit of the continuum, including reflection.}
\begin{center}
\begin{tabular}{lll}
\hline\hline

$T_{\rm mbb}$ & keV &  $0.21\pm0.3$\\
$\Gamma$ & &  $2.5\pm0.1$\\
{\bf blurred} &&\\
$norm_{\rm pow}$ & $10^{51}$\,ph\,s$^{-1}$\,keV$^{-1}$& $4.1\pm0.3$\\
$\Gamma$ &  & $1.73\pm0.08$\\
$\alpha$ & & $>1.7$\\
$i$ & deg &  $<30$\\
$s$&& $0.8\pm0.3$\\
{\bf unblurred}&&\\
$norm_{\rm pow}$ & $10^{51}$\,ph\,s$^{-1}$\,keV$^{-1}$&  $8.0\pm2.5$\\
$\Gamma$ &&  $1.73\pm0.08$\\
$i$ & deg &  60 fixed\\
$s$&&  $1.1^{+1.3}_{-0.2}$\\
$\chi^2/\nu$ & & 322/230\\
\hline

\hline
\end{tabular}
\end{center}
Note: The model includes a black body, a power law, blurred 
reflection from the accretion disk and reflection from unblurred cold and distant material.  
The value of $\chi^2_{\rm red}$ refers to full modeling, including absorption (Sect.~\ref{par:abs})
and emission lines (Sect.~\ref{par:broad}).
\end{table}
\subsection{The contribution of the BLR to the \fek\ line}
We have quantitatively evaluated the BLR contribution to the emission of the prominent 
\fek\ line at 6.4\,keV. The highly ionized specie of \fexxvi\ has a negligible role in the BLR, 
therefore the emission we see at those
energies in the X-ray spectrum must come from some other location (Sect.~\ref{par:fexxvi}).  
We based this analysis on previous results for the BLR structure and on the present detection of the \ovii\ broad line in the
RGS spectrum. From the extensive experience of broad lines studies in optical spectra of AGN, 
it is known that measurements of 
broad and shallow emission features can be affected by large uncertainties and should be cautiously treated. 
The limitation is due for instance 
to the signal-to-noise ratio of the continuum level. Resolution, effective area and calibration of the instrument are
also relevant effects. However, known, possibly not completely calibrated, instrumental narrow features \citep[23.05,
23.35\AA,][]{cor03} fall at the border of the \ovii\ emission region. Observation-specific bad pixels (flagged as bad
channels in the spectrum) are automatically 
taken into account when computing the instrument response, but they may certainly cause additional uncertainty 
when fitting narrow features. The \ovii\ broad line is 1.4\AA\ wide and the determination of its flux is not
significantly influenced by bad pixels. Considering also the very significant, independent, detection in the \ch-LETGS
data, we treat this feature as an intrinsically-low-flux \ovii\ line. The line flux measured in 2005 is indeed weaker than in 2003 (Sect.~\ref{par:broad}), 
even though the SED of 2005 shows an increased availability of optical and UV photons (Fig.~\ref{f:sed}). 
Such a behavior of the line intensity is however not
unexpected, as the single line luminosities are also sensitive to the long term flux history and the 
shape of the SED both in the UV and X-ray band (at least of the previous two
months, given the typical size of the BLR). The X-ray continuum of \mrk\ has been observed to change in a 
month time-scale (Sect.~\ref{par:var}), however the information on the light curve of \mrk\ is sparse, as the most
recent observation prior to the \xmm\ pointing has been done in 2003 (C07). Therefore a reconstruction of the SED prior to the
present \xmm\ observation cannot be performed.
In this study we find that only a small fraction of the broad \fek\ line can be produced in the BLR. 
From the LOC modeling we find that the
\fek\ line luminosity, coming from the BLR, should be about six times smaller than the \ovii\ triplet. 
This translates in a 
contribution to the \fek\ line which is about 30 times smaller than that observed, implying a 3\% contribution. 
However, the elemental abundance close to a
black hole, especially of iron,
might be different from Solar. Abundances up to 7 times Solar have been suggested \citep[e.g.][]{fabian02}. 
The same BLR model,
but with the Fe abundance enhanced by a factor seven only reduces the \ovii/\fek\ luminosity ratio to about 4.5. 
However, such an overabundance implies much stronger X-ray iron lines from the L--M shells, 
which are not observed. If we artificially consider the luminosity of the \fek\ as entirely produced in the BLR, we should observe an \ovii\
line with a luminosity as large as $3\times10^{42}$\,erg\,s$^{-1}$, which is clearly inconsistent with both the present and
archival measurements (C07). Moreover the \neix\ and \oviii\ broad lines (with predicted luminosity of about 
$60\times10^{40}$\,erg\,s$^{-1}$) would be clearly visible in the spectrum, against the observational evidence.
The same discrepancy applies if we consider the very broad profile of the \fek\ line as arising from the BLR
(Sect.~\ref{par:broad}). In that case, 
the BLR contribution would be 23 times smaller than observed. The unresolved component of the \fek\ would instead be
about 6 times smaller. If this were the case, the contribution of the BLR to the \fek\ line would not be negligible
($\sim16$\%). However, the FWHM of the core of the \fek\ line, as measured by \ch-HETGS is $4200^{+3350}_{-2950}$\,\kms
\citep{nandra06}, which is inconsistent with the FWHM of the UV lines \citep{gabel05}. Moreover, BLR line are subjected to
significant variability both in the UV \citep[e.g.,][]{gk98} and in the X-rays \citep[][and present work]{steen09}, 
while we could not detect any significant change in the iron line flux, 
measured in a few occasions by different instruments (Sect.~\ref{par:var}). 
This also supports the idea that the BLR contribution should be minimal. 
An important note is that the BLR parameters of \mrk\ were determined first using UV data only 
(see C07 and Sect.~\ref{par:loc}) and then applied to the X-ray data. With this approach, any 
X-ray line only provides useful
upper limits, within the range imposed already by the $C_V$ value, on the normalization of the LOC model
(Fig.~\ref{f:loc_gamma}). Therefore, in principle, given a BLR model and an SED, the contribution of the 
BLR to the \fek\ line can be calculated without the support of other X-ray lines. 
As a test, we also predicted the iron line
luminosity considering the SED and BLR parameters derived for the 2003 data set. In that case,
reminding that we do not have any simultaneous measurement of the \fek\ line in 2003, the contribution of the BLR to the
line would be roughly 17\%. 

Here for the first time we have tested the connection between the optical BLR and the \fek\ line, using the 
intrinsic luminosity of the lines (rather than the FWHM) 
and relying on a possible physical model for the BLR \citep[The LOC model,][]{baldwin95}. 
The idea of a non-relation between the BLR and
the \fek\ line has been already tested comparing the width of the optical 
broad lines with the width of \fek\ line \citep[e.g.][]{sulentic98,nandra06}. In at least one case, 
\citep[the LINER galaxy NGC~7213,][]{bianchi08} a simultaneous optical--X-ray observation revealed the same 
width for the H$\alpha$ and \fek\ lines,
suggesting a common origin in the BLR. However the UV and X-ray lines might come in regions closer to the black
hole with respect to the lower ionization lines in the optical, implying a larger FWHM. 
Moreover NGC~7213 can be considered an outsider object, as has been proven to totally lack the reflection component
\citep{bianchi03}. Therefore a
direct comparison with classical Seyfert~1 galaxies (as is \mrk) may not be possible.   
\subsection{The origin of the \fek\ line}\label{par:origin}
Any variability of the iron line would easily suggest a close 
interaction between the central source and the region producing the
iron line, e.g. the accretion disk. The line does not need to be relativistically smeared, as 
also Gaussian-shaped lines may arise from the accretion disk \citep{yaqoob03}.  
Using ASCA data \citet{weaver01} 
found variability in the line centroid at the 1.6$\sigma$ level (90\% confidence).
The variability was observed on a time scale of about 20\,ks following a 2--10\,keV central-flux variation of $\sim$15\%.    
The average EW and FWHM as measured by ASCA were however consistent with the present \xmm\ data. 
Over a time scale of at least 3 years, the \fek\ line are, in first approximation, 
stable in luminosity, width and centroid energy 
within the errors (Table~\ref{t:fe_var}). To which extent the line is stable is difficult to assess, as some measurements are
affected by large uncertainties. Formally, the line could have changed of as much as a factor of two or more 
in luminosity. In a scenario where a dominant component of the line does not respond to the 
central source variation, 
an origin of the iron line in the accretion disk can be justified by light-bending \citep[e.g.][]{fv03,luigi07}.  
In that case the apparent flux is allowed to
change even of a large factor (up to four)  while the line-flux change is marginal \citep{giovanni03}. 
A number of factors may influence the iron line behavior. The limited knowledge of the exact nature of the 
emission components (both \fek\ line and continuum) in this source does not allow us to 
make predictions on the variation
of the line in response to the continuum changes on either short or long time scales.\\ 
A symmetric line profile, possibly constant over a long time scale, may also suggest a
relatively unperturbed environment, like for example the molecular torus, $\sim$1\,pc away from the central source. 
In the
framework of the unification model \citep[e.g.][]{antonucci}, the iron line is 
a natural consequence of the obscuring torus.
In type~1 objects, the expected equivalent width would be of the order of 100\,eV. 
However geometrical effects would easily
reduce it of a factor of two \citep{nandra07,kk87}. In the case of \mrk, the EW of the narrow iron line 
is $70\pm15$\,eV, in agreement with the
theoretical prediction. The presence of Fe\,K\,$\beta$, insures that the ionization state of iron is less than \fexvii\
\citep{yaqoob07}. A further constraint on the ionization state of iron would be based on the \fek/Fe\,K\,$\beta$ branching
ratio, which ranges from 0.12 for \fei\ up to 0.17 for \feix\ \citep{palmeri}. Unfortunately, considering 
the associated errors on the observed line fluxes, we only obtain a lower limit of 0.1 to that ratio. 
\subsection{The nature of the \fexxvi\ line}\label{par:fexxvi}
He-like, sometimes associated with H-like, iron lines have been detected in a number of 
AGN spectra. These lines can be associated to high column density, 
photoionized gas \citep[][and references therein]{annalia,bianchi05}. 
The spectrum of \mrk\ does exhibit a narrow emission line at $\sim$6.9\,keV, consistent with emission from \fexxvi, but 
interestingly \fexxv\ is not detected, showing the presence of extremely highly ionized gas. 
Interpreting this emission as a product of photoionization, we find that a high column-density ($N_{\rm H}\sim10^{23}$\,cm$^{-2}$) with a 
very-high ionization parameter (log$\xi\sim 6$) is necessary. A range of $N_{\rm H}$ and log$\xi$ and covering factors 
is allowed to explain the \fexxvi\ emission and at the same time be
consistent with the measured upper limits on the other high ionization lines: \fexxv, \nex\ and \oviii\ (see
Sect.~\ref{par:high} and Fig.~\ref{f:high_xi_cv}). The covering factor is on average 0.65. 
As we do not see any associated absorption, this gas must be
out of our line of sight. The EW of the \fexxvi\ line ($12\pm5$\,eV) is consistent with an origin 
in photoionized circumnuclear gas at the distance of the obscuring torus \citep{bianchi_matt}. The torus
itself would produce the bulk of the neutral-weakly ionized iron emission, while a highly
ionized outer layer of photoionized gas would be responsible for H-like and He-like iron. In our case the gas is
so ionized to
suppress even \fexxv.\\
Another possibility is that the \fexxvi\ line is formed in a hot galactic wind, driven by starburst activity
\citep{che_clegg85}. However, a very hot (log~E$\sim$9--10\,keV) plasma would be needed to produce the observed \fexxvi\ line
luminosity via collisional ionization. Such high temperatures have been proposed for example for the diffuse
emission of the Galactic ridge \citep[e.g.][]{tanaka02}. However, this interpretation is problematic with respect to 
the confinement of the plasma within the host galaxy, the source of such heat \citep[e.g.][]{masai02} and the
contamination by point-like, hard sources in the diffuse emission spectrum \citep[e.g. in \object{NGC~253},][]{strickland02}. 
Another way to model the high ionization spectrum in terms of a galactic wind, is to invoke quasi-thermal
acceleration of electrons which then would produce X-rays with thermal photons. Iron lines would be produced by
recombination of background iron ions recapturing electrons, as proposed by \citet{masai02} for the Galactic ridge
emission.   

\section{Conclusions}
We have performed a detailed modeling of the spectrum of \mrk, observed for $\sim160$\,ks by \xmm. Thanks to the broad band coverage we had the opportunity to study in detail
both the low-energy emission lines (observed by RGS) and the iron K complex at around 6.4\,keV (observed by Epic).
 
We have extended the ``locally optimally emitting cloud" model, which has been conceived to model the UV broad lines coming
from the BLR \citep{baldwin95}, first
to the soft X-ray band (C07) and, in the present paper, to the \fek\ line. 
The BLR can easily account for
the \ovii\ broad emission (C07), while it contributes marginally to
the luminosity of the broad iron line at 6.4\,keV (about 3\%, this work). 
This is the first attempt to evaluate the BLR contribution to the luminosity of the iron line using a global modeling 
(rather than relying on the line FWHM only). Further investigation
 on high quality data of similar sources are of course necessary to test the robustness of our results.\\
   
The bulk of the \fek\ narrow emission line shows no remarkable variability, 
in flux and shape over at least three years (Sect.~\ref{par:var}) and it may be produced by reflection 
from distant and cold matter. 
The EW of the line ($\sim$70\,eV) is also consistent with this picture.\\
The very broad component of the \fek\ line may be modeled by a line arising from an accretion disk, 
associated with relativistically 
smeared reflection. However, a formal fit using broad-band reflection models is affected by large 
uncertainties and degeneracy of the
parameters, as we cannot access
the region above 10\,keV and we only have to rely on the \fek\ profile, 
which is in turn formed by two line-components.\\

We also detected emission from \fexxvi, but not from \fexxv. 
This implies an extremely highly ionized medium. Using the
available constraints on other highly ionized ions (e.g. \nex, \oviii\ and \nvii) we could limit the column density of such gas
$10^{22}<N_{\rm H}<5\times10^{23}$\,cm$^{-2}$ with an increasingly higher ionization parameter $5.3<{\rm log}\xi<7$. The
covering factor of the gas is 0.65 on average. Such a
gas is predicted by the AGN unification model, and it could be associated with a highly ionized outer layer of the 
obscuring torus \citep[e.g.][]{kk87}.\\ 

\begin{acknowledgements}
The Space Research Organization of the Netherlands is supported
financially by NWO, the Netherlands Organization for Scientific
Research. 
\end{acknowledgements}


\end{document}